\documentclass[preprint,aps]{revtex4}

\usepackage{graphicx}
\usepackage{dcolumn}
\usepackage{bm}

\begin{document}

\preprint{1dwire}

\title{Quantum Size Effect and Electronic Stability of Freestanding Metal Atom Wires}

\author{Haiping Lan}
\author{Ping Cui}
 \affiliation{ICQD, University of Science and Technology of China, Hefei, Anhui, 230026, China}
\author{Jun-Hyung Cho}%
\affiliation{%
Department of Physics and Research Institute for Natural Sciences, Hanyang University, 17 Haengdang-Dong, Seongdong-Ku, Seoul 133-791, Korea
}%
\author{Qian Niu}
\affiliation{
Department of Physics, The University of Texas at Austin, Austin, Texas 78712,USA}%
\author{Jinlong Yang}
\affiliation{
ICQD, University of Science and Technology of China, Hefei, Anhui, 230026, China
}
\author{Zhenyu Zhang}
\affiliation{
ICQD, University of Science and Technology of China, Hefei, Anhui, 230026, China
}
\affiliation{Department of Physics and Astronomy, University of Tennessee, Knoxville, Tennessee 37996, USA}
\date{\today}

\begin{abstract}
Fabrication of freestanding or supported metal atom wires may 
offer unprecedented opportunities for investigating exotic behaviors 
of one-dimensional systems, including the possible existence of 
non-Fermi liquids. Many recent efforts have been devoted to 
the formation of different kinds of metal atom wires in 
freestanding forms by novel techniques like mechanical break junction 
or deposited on substrates via self-assembly, 
focusing on their mechanical, chemical and electronic properties. 
Various atom wires with different lengths can be obtained during fabricating processes. 
Their size distributions  have been extensively analyzed, 
which exhibit diverse features. 
Although several factors such as strain and substrate effects have been 
employed to interpret these phenomena, 
the stability of atom wire itself is largely ignored. 
Using density functional theory calculations, 
we present a thorough study on freestanding metal atom wires, 
including \textit{s}, \textit{sd} and \textit{sp} electron prototypes, 
to examine the size effect in their stabilities. 
We find that the total energy of all systems oscillates 
within wire length, which clearly indicates the existence of some preferred lengths.
Increasing the length of  atom wires, \textit{s} electron system shows even-odd 
oscillation following a $a/x+b/x^2$  trend in the stability, 
due to both electrons pairing up and one-dimensional quantum confinement. 
Meanwhile, \textit{sd} electron systems show a similar oscillation 
within wire length although \textit{s-d} hybridization is presented. 
In \textit{sp} electron systems, some oscillations beyond the even-odd one are 
exhibited due to unpaired \textit{p} orbitals 
resulting  in some nontrival filling rule. 
Our findings clearly demonstrate that electronic contribution is 
quite critical to the stability of freestanding atom wires 
and is also expected to dominate even 
when atom wires are deposited on substrates or under strain. 
This study sheds light on the formation of metal atom wires and helps understanding relevant phenomena.	
\end{abstract}

\pacs{Valid PACS appear here}
\maketitle

\section{INTRODUCTION}
As a model of one-dimensional (1-D) systems, metal atom wires 
have attracted enormous attentions to investigate relevant exotic 1-D behaviors, including the existence of non-Fermi liquids\cite{snijders10,springborg07}. 
Various experimental techniques have been developed to fabricate different kinds of metal atom wires. 
There have been many reports that atom wires can be self-assembled 
or manipulated on semiconducting or metallic substrates, such as Au/Si(111), Au/Si(557),Ag/Si(55 12), Pb/Si(557), Ga/Si(100) and In/Si(111)\cite{snijders10,springborg07,erwin98,segovia99,kim07,ahn02,robinson02,snijders05,gonzale04,nilius02}.
Furthermore, freestanding atom wires have also been obtained via some novel methods 
such a mechanical break junctions (MBJ)\cite{yanson98,springborg07}. 
All these progresses greatly spark diverse investigations 
on properties/behaviors of metal atom wires. 
For example, Segovia \textit{et al.} measured band structure of Au/Si(557) 
by angle-resolved photoemission spectroscopy(ARPES), 
and suggested the existence of a Luttinger liquid\cite{segovia99}. 
A later work on temperature-dependent ARPES and scanning tunnel microscopy 
study  showed there is a symmetry-breaking metal transition in 
Au/Si(557) which can be interpreted as a traditional Peierls transition, 
precluding the formation of a Luttinger liquid at low temperature\cite{ahn03}. 
In addition, a recent measurement by electron energy loss spectroscopy revealed significant dynamic exchange correlation effects on the 1D plasmon 
despite its high electron density and large Fermi velocity\cite{nagao06}. 
Using MBJ technique, Yanson \textit{et al.} explored Au atom wires, and found a wire with 7-atom length can be formed 
which behaves as a perfectly quantized one-dimensional conductor\cite{yanson98}.  
All these investigations clearly demonstrate that metal atom wires are desirable workhorses to  testify theoretic predictions of 1-D systems.

To explore these exotic properties of 1-D systems, 
longer or defect-free metal atom wires are quite critical. 
Therefore, understanding the formation mechanism of these wires should help 
to improve our ability to control the fabrications such as 
formation wires by different materials and longer ones. 
Smit \textit{et al.} gave a systematic study over 
5\textit{d} metals by employing MBJ, and suggested a stronger bonding of 
low-coordination atoms with respect to 4\textit{d} metals 
is due to \textit{sd} competition caused by the relativistic effects\cite{smit01}. 
Moreover, sequential theoretical simulations confirmed that 5\textit{d} metal atom wires 
like Au and Pt have bonds much stronger than their bulk ones\cite{bahn01}.
Given supported atom wires, a highly anisotropic substrate 
as the growth template is very important. Usually, surfaces with a periodic 
step structures like vicinal metal or semiconductor surfaces are 
employed in molecular beam epitaxy to fabricate arrays of atom wires. 
Once adatoms diffuse anisotropically on these surfaces, 
well-ordered 1-D atom wires are possibly self-assembled\cite{javorsky09,oncel08}.

Strictly spearking, there are no  ideal 1-D systems like infinite atom wires 
exist in experiments but segments of different lengths obtained 
in fabrications. Such wires of varying lengths have been observed in 
some heteroepitaxial self-assembled systems and MBJ experiments\cite{snijders10,yanson98}. 
The length distribution can affect the phase transition temperature, change the conductivity, 
inhibit charge orders, and alter the effective dimensionality of the system. 
Thus, it is essential to understand the factors leading to different sizes 
and to control the wire lengths. 
So far, a few works have suggested the length distribution generally depends on 
several factors including the deposition coverage, 
the substrate temperature, adatoms adsorption energy and surface defects\cite{albao05,gambardella06,stinchcombe08,javorsky09,tokar07,tokar03,kocan07}.
With respect to different metal atom wires, 
the length distributions showed quite different behaviors. 
Some experimental works reported that the length distributions of Ga and In atom 
wires on Si(100) surface monotonously decrease for various coverages, 
and have certain scale relations\cite{albao05,javorsky09}. 
Comparing with experimental data of Ag wires on Pt(997) surface, 
Gambardella \textit{et al.} examined the size distribution of atom wires 
in the framework of 1-D lattice gas model, 
and found that the lengths obey the geometric distribution\cite{gambardella06}. 
This length distribution indicated that an atom binding to the wire is independent of its length, 
thus suggesting  only nearest-interactions 
account for its growth. A later work by Tokar \textit{et al.} showed that
incorporating additional interactions like elastic strain and charge transfer 
can obtain higher accurate fitting results\cite{tokar07}. 
Whereas, an extensive analysis based on STM by Crain \textit{et al.} found
an oscillating length distribution is exhibited in Au/Si(553)\cite{crain06}. 
They ascertained short-range interactions like local rebonding of the surface 
result of a strong peak at a length of one atom, and found that even wire lengths are 
favored over odd lengths up to lengths of at least 16 atoms, 
which indicates the quantum size effect plays an important role. 
This growth feature is in analogy to the  electron growth of thin film\cite{zz98},
which implies pure electronic contribution can exert a fundmental impact on wires' growth. 
Further theoretical analysis by Souza \textit{et al.} suggested 
a model only within electronic structure can capture the feature of 
the even-odd oscillation, and qualitatively agreed with the experimental 
length distribution\cite{souza08}.

As revealed in both experiments and preliminary tight-binding calculations, 
the quantum size effect can dominate Au atom wire's growth\cite{crain06,souza08}, 
and result of an even-odd oscillation. Except Au system, few  attentions  
have been given for the other metals  about this aspect. 
Hence, it should be interesting to explore  how other metal atom wires 
are modulated by the quantum size effect and some relevant 
impacts on their stabilites. 
Our purpose  is to present a systematic \textit{ab initio} study 
on the quantum size effect in linear metal atom wires' stability,
with a special focus on its development with wire length.
We choose several prototypes of \textit{s}, \textit{sd} and \textit{sp} 
electron systems such as Na, Ag, Au, Ga, In and Pb atom wires 
in freestanding form for these atom wires all have been 
obtained experimentally\cite{snijders10,springborg07}. 
We hope our study will facilitate the understandings of
various size distributions even under the influence of substrates, 
strain and charge transfer. 
We find that the total energies of all systems oscillate within wire length, 
which clearly indicates the existence of some preferred lengths. 
In particular, \textit{s} electron system shows an even-odd 
oscillation following a $a/x+b/x^2$ trend in the stability when the length is increased, 
which results from  the pair-up of electrons and 1-D quantum confinement along the wire. 
Meanwhile, \textit{sd} electron systems show a similar oscillation 
with the wire length although \textit{s-d} hybridization occurs. 
In \textit{sp} electron systems, 
some beyond even-odd oscillations are exhibited as unpaired \textit{p} 
orbitals result in some nontrival filling rule, 
which almost washes out the effect of 1-D confinement because of their localized bonding characteristics.

The rest of this paper is organized as follows. 
The computational methods and numerical  details are presented in Sec. II. 
Sec. III gives  the results of various metal atom wires. 
The quantum size effect and electronic stability  are studied in detail. 
A brief summary  is then given  in Sec IV.

\section{CALCULATION METHODS}
All results we present here were carried out by the projector augmented-wave(PAW)
method\cite{paw} implemented in the Vienna \textit{ab initio} simulation package(VASP)\cite{vasp1,vasp2}. 
Based on density-functional theory(DFT), exchange and correlation effects were 
described by the generalized gradient approximation(GGA-PBE)\cite{pbe}. 
The energy cutoff for plane-wave basis of 300.0 eV was used for all examined 
systems, which ensured the total energy converging into 0.01 eV per atom. 
In particular, we employed a small Gaussian broadening width 0.01 eV to 
achieve integer occupation  of states.
Isolated atom wire was simulated in a rectangular box with periodic 
images' separation larger than 10 ~\AA~, 
and the longest one was up to 20 atoms for each metal. 
As for relevant infinite systems, we sampled 1-D Brillouin zone within 40 Monkhorst-Pack grids
after carefully checking the covergence of k-point sampling. 
Here, we adopted one atom per unitcell without explicit  consideration of Peierls effect
since further checks on the two-atom unitcell neither showed any apparent dimerization
nor made any difference in bandstructure but just folded up the 1-D Brillouin zone.
All geometric structures were relaxed using the conjugate gradient method until residual force per atom was less than 0.01 eV/~\AA~.

Since 1-D atom wires are likely to be metastable for all involved metals, 
here we adopted initial configurations for atom wires by 
setting bond lengths to that of the infinite one. 
Because of certain fine interactions resulting from magnetic coupling for 
\textit{sd} and \textit{sp} electron systems, 
it is impossible for us to conduct a thorough study of
various combinations over initial magnetic moments of atoms to obtain a local minimum, 
especially for a wire of tens of atoms. 
Although there are some ambiguities in energy minimums of atom wires,
these fine interactions are unlikely to yield major changes or 
affect dominant electronic interactions between atoms.
So we used several simple combinations within initial magnetic moments 
of atoms like  $\uparrow\uparrow\uparrow\uparrow\uparrow\uparrow$(ferromagnetic),
$\uparrow\downarrow\uparrow\downarrow\uparrow\downarrow$(antiferromagnetic),
$\uparrow\uparrow\uparrow\downarrow\downarrow\downarrow$(antiferromagnetic),
$\uparrow\uparrow\downarrow\downarrow\uparrow\uparrow$,etc. for 6-atom wire 
in further geometric relaxation, and chose the energy minimum among them. 
A similar setting for initial conditions of different
magnetic moments has been discussed in previous work 
on transition metal atom wires while examining average cohesive energy of various lengths\cite{atca08}.
Our calculation results showed  energy uncertainty  of the minimum is less than 0.1 eV (about 0.01 eV per atom), ensuring us to reach further conclusions.

To characterize the electronic stability of atom wires and the corresponding size effect, 
we use the cohesive energy\cite{zz98,cho98},reading
   \begin{eqnarray}
 E_c(n)=(E_t(n)-n\cdot E_t(1))/n,
	   \label{eq1}
   \end{eqnarray}
to examine  atom wires' stability in different length, where $E_t(n)$ is the total energy of the
atom wire of $n$ atoms.  Thus, the larger the absolute value of cohesive energy  is, 
the more stable the wire segment will be. 
This energy value purely reflects the contribution of electronic binding interaction between atoms.
We then introduce the second difference of $E_c$,reading
\begin{eqnarray}
	d^2E(n)=E_c(n+1)+E_c(n-1)-2\cdot E_c(n),
	\label{eq2}
\end{eqnarray}
as the criterion for the stability of a $sp$ metal wire segment 
in analogy with that for a thin film\cite{zz98}. 
According to Eq.(\ref{eq2}), 
a wire of $n$ atoms is stable when $d^2E(n)>0$ and unstable otherwise.

\section{RESULTS AND DISCUSSIONS}

\subsection{Na atom wires}
Firstly, we consider Na atom wires.  
So far, considerable attention in Na wire has been drawn to its electron transport properties, especially 
to its  quantum conductance measurements\cite{springborg07}. 
Yanson \textit{et al.} 
experimentally studied the conductance through Na thin junction via MBJ technique, 
and found the conductance histogram shows a typical oscillation on measured times due to the
shell effect, a pure electronic effect  resulting from size confinement\cite{yanson99}. 
Further theoretical works were then extended to  situations of atom wire\cite{barnett97,lee04,havu02}, 
and calculated conductance showed an oscillation with wire length, which also largely depended on 
the structure of nanocontacts\cite{sim01,havu02}. 

\begin{figure}
	\includegraphics[width=0.64\textwidth]{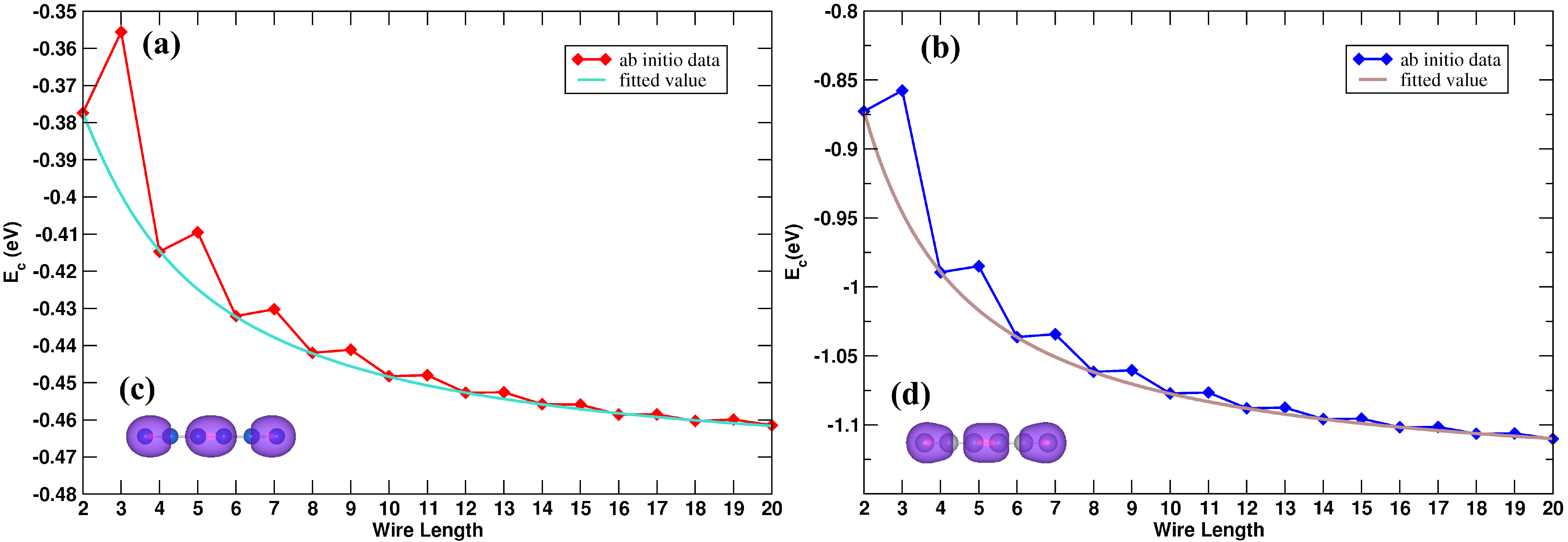}%
	\caption{(Color online)\label{fig01} The cohesive energy  $E_c$ of (a) Na atom wires, and (b) Ag atom wires versus lengths. Fitting data  are also shown in (a) and (b), respectively. 
HOWSs of (c) 6-atom Na wire and (d) 6-atom Ag wire are displayed, respectively.  }
\end{figure}

Since sodium has only one valence, 
the stability of atom wires would be greatly enhanced once electrons pair up. 
The cohesive energy therefore shows an even-odd oscillation with 
the length, as shown in Fig.\ref{fig01}(a).  
It is clear that the pair-up of electrons is much more pronounced for shorter wires, 
and gains small energy less than 0.01 eV when the length is up to 10-atom or more.
Thus, relevant even-odd oscillation  would be smeared off for longer wires, 
which suggests the quantum size effect can not be easily observed in the experiment 
as a bit strain  or charge transfer is supposed to
wash out the  related oscillation .
Actually, 
the cohesive energy converges in a $a/x +b/x^2$ trend as the length increases, 
here $x$ is the wire length, and $a$,$b$ are constants.  
This trend is simply caused by the quantum confinement along the wire direction, 
since the energy level $E_i$ of 
a 1-D square quantum well is inversely proportional to the square of its width $W$ ,i.e,
$E_i\sim i^2/W^2$, as already shown in a particle-in-a-box model. As a result, the total energy 
of $n$ electrons is the sum over the energy levels and proportional to $n(n+1)(2n+1)/n^2$ given $W\sim n$.
Therefore, the cohesive energy per electron follows the $a/x + b/x^2$ trend in this simple model.
This remarkable agreement between \textit{ab initio} results and the particle-in-a-box model is clearly due to
the delocalized and undirectional dependence of the \textit{s} orbitals of Na atom wires.
The highest occupied wire state (HOWS) for 6-atom wire is then presented in Fig.\ref{fig01}(c),
showing a typical $ss\sigma$ binding character with two nodes along wire.  
Clearly, the number of nodes in HOWS depends on  the wire length, 
with $(n-1)/2$ for odd one while $(n-2)/2$  for even situation,  stemming from  
the orthogonality relations between the wavefunctions. 


\subsection{Ag and Au atom wires}
Recently, both Ag and Au atom wires were fabiracated experimentally.
It was reported that ultrathin silver wires with a width 0.4 nm  
were successfully synthesized inside  nanotubes\cite{hong01}. 
Some MBJ techniques and atom manipulation by STM were  extended to 
both Ag and Au systems for fabricating nanowires\cite{smit01,yanson98,rodrigues02,sperl08,nilius02}. 
In this case, the structure of nanowire is likely to be a linear atom wire.  
In addition, 
some experimental efforts have been devoted to depositing Ag atom wire in Si vicinal surface\cite{ahn02}.
All these experimental progresses encouraged extensive theoretic studies on various aspects of 
Ag atom wires, especially on electric properties and quantum transport\cite{springborg07,springborg03,riberio03}.   With regard to Au atom wire, there have been enormous works exploring its various properties since its first experimental fabrication\cite{yanson98,ohnishi98,snijders10,springborg07}. 
A few attempts have been  devoted  to the study of its transport properties by MBJ techniques, 
and showed Au atom wire can survive 
in a longer length in comparison with Ag system due to 
stronger \textit{sd} competition\cite{yanson98,smit01}. Many efforts have also
focused on  possible realizations of Luttinger liquids and 
other 1-D exotic behaviors
by depositing Au atom wires in various vicinal surfaces 
such as Si(553), Si(557), Si(337), Ge(100)\cite{snijders10,crain06,robinson02,crain05}.
All these works boost several seminal explorations such as end modes of 1-D systems and atom wires' electronic growth\cite{crain05,yan07,crain06}.   
And experimentally,
both Ag and Au atom wires showed size effects can largely modulate their electronic properties and stability\cite{sperl08,crain06}.

Silver is of the closed 4\textit{d} shell, behaving as a single valence system. 
 As displayed in Fig.\ref{fig01}(b), the cohesive energy of Ag atom wires shows 
 a similar even-odd oscillation  as
that of Na atom wires, and relevant values are around -1.0 eV.
Since certain \textit{sd} hybridization is likely to  enhance the binding interaction 
between atoms, Ag atom wires show a bit larger amplitude of the even-odd oscillation 
in comparison with Na atom wires. This enhanced interaction should
lead to observing relevant quantum size phenomena in real experiments. 
Acutally, recent STM measurements have revealed that resonances of unoccupied states show 
a strong size dependence of energy values on Ag/Ag(111) system\cite{sperl08}, 
which suggestes the size effect survives even under the influence of substrate.
Therefore, we can expect relevant quantum size effects would exhibit in 
some self-assembled Ag wires.
And we  present HOWS for 6-atom wire in Fig.\ref{fig01}(d), 
which clearly shows a bit different character from that of Na atom wire due to some \textit{sd} hybridization.  Meanwhile, the number of nodes of HOWS is also the same as that of Na systems.

\begin{figure}
	\includegraphics[width=0.96\textwidth]{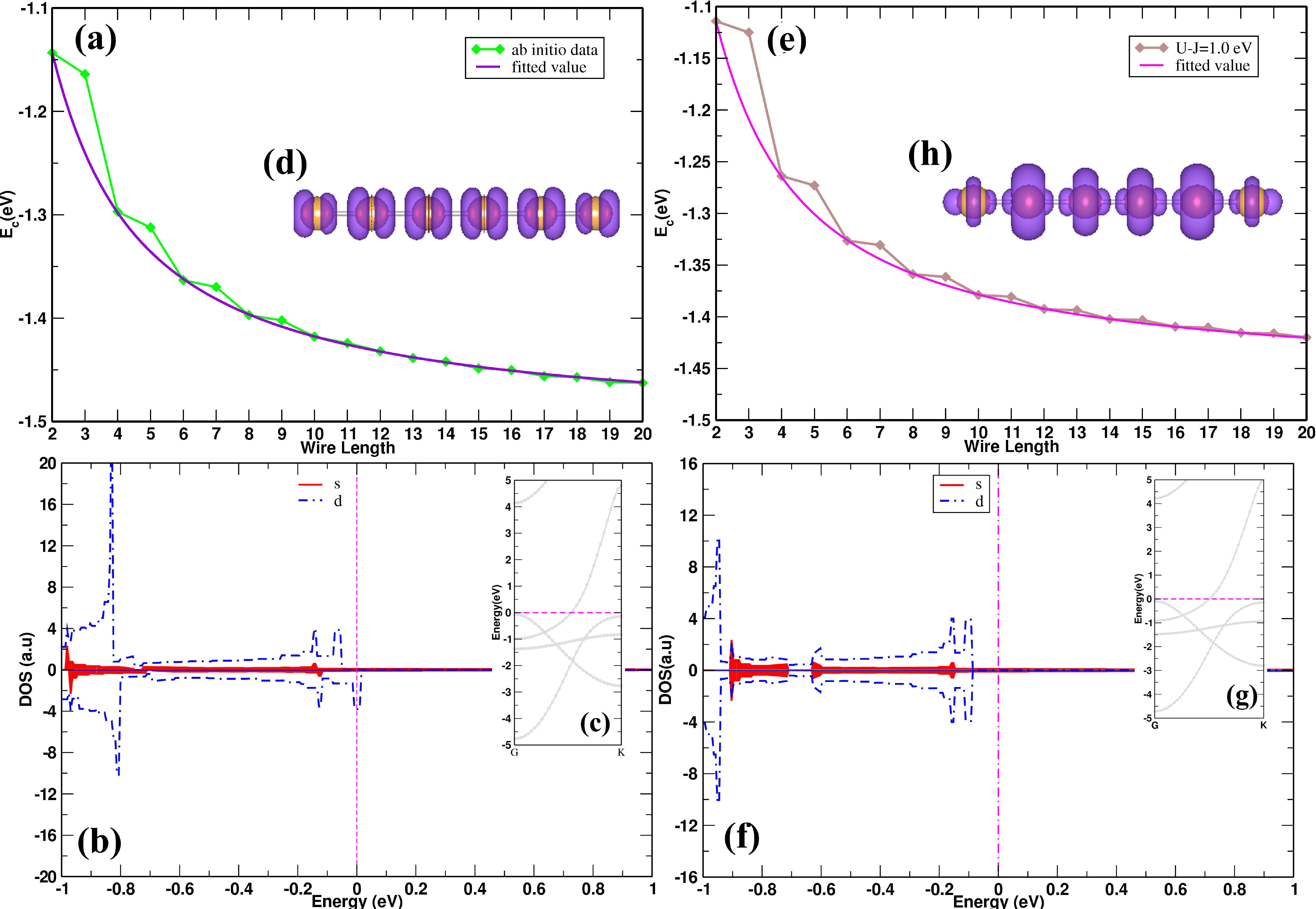}%
	\caption{(Color online)\label{fig02} (a) The cohesive energy $E_c$ of Au atom wires versus length, (b) and (c) are density of states(DOS) and bandstructure of the infinite Au wire, respectively.
(e) The cohesive energy $E_c$ of Au atom wires versus length by GGA+U calculations within $U_{eff}=U-J=1.0$ eV,(f) and (g) are corresponding to DOS and bandstructure of the infinite Au wire. HOWS of 6-atom is then given in (h). In both cases, the Fermi level is the zero energy. Fitting data  is also shown in (e). }
\end{figure}

In contrast to Ag, relativistic effects are quite large in Au system, 
which result in strong \textit{sd} competition.
Consequently, the \textit{s} shell is contracted and the \textit{d} electrons move up in comparison with Ag. 
This electronic feature leads to stronger binding interaction in Au 
wires, and the cohesive energy approaches -1.5 eV when the length is increased 
as shown in Fig.\ref{fig02}(a).  
When the length is shorter than 13 atoms, the cohesive energy shows almost 
the same behavior as that of the Na and Ag wires.  However, a crossover occurs when $n=13$, 
leading to a transition of the oscillation. 
This behavior can be interpreted by 
the bandstructure of infinite wire as shown in Fig.\ref{fig02}(b) and (c). 
Due to the axial symmetry, \textit{d} bands split into 3 branches,~e.g, $d_{z^2}$, $d_{x^2-y^2}$/$d_{xy}$ 
and $d_{xz}$/$d_{yz}$ . 
Then, four band branches below the Fermi level are shown in Fig.\ref{fig02}(c), respectively. 
At $\Gamma$ point, the lowest band branch is mostly of the $d_{z}$ character, which then 
develops across 1-D Brillouin zone by approaching the Fermi level at $K$ point. 
However, $d_{x^2-y^2}$/$d_{xy}$ branch is almost a flat band across 1-D Brillouin zone.
This dispersionless character suggests there are no bonding interaction between the $d_{x^2-y^2}$/$d_{xy}$ electrons of atoms and they are just localized around Au atom itself.   
The next band branch, crossing the Fermi level at the middle of 1-D Brillouin zone,
is mostly of the $s$ character.
The last branch, $d_{xz}$/$d_{yz}$,
is then in the vicinity of the Fermi level at $\Gamma$ point.
At $\Gamma$ and $K$, two special symmetric points of 1-D systems, the band dispersions are all 
horizontal, leading to very sharp van Hove singularities of 1-D system 
which result in sharp peaks of density of states(DOS) as displayed in Fig.\ref{fig02}(b).
Since the bands  mostly have \textit{d} character at the edges,
the exchange energy gain could be rather large when a band spin splits so
that one of the spinchannels band edges ups above the Fermi level, 
and the other then downs below. 
Thus, if a band edge ends up sufficiently near the Fermi level, 
e.g., appearance of sharp peaks of DOS,
we can predict a magnetic moment \cite{delin03},
which is known as the Stoner instability.
Meanwhile, this instability depends closely on Au bond length, and 
a bit elongating(shortening) bond can eliminate(enhance) the related ferromagnetic behavior
as also reported in previous works\cite{delin03}. 
Such behaviors mean the bond lengths of Au wires will affect significantly 
the positions of band edges .
As for finite wires of different lengths, bonds oscillate around that of infinite one.
Correspondingly, their HOWSs show different characters.
When the wire length is less than 4-atom, bonds are smaller than that of infinite wire (2.59~\AA), 
 and HOWSs are of the $d_{z^2}$ character, partially with the $s$ contribution.
4-atom wire is of two bonds less than 2.59~\AA, large part of HOWS also shows  $d_{z^2}$ character.
Longer than  the number of 4-atom, central bonds of atom wires are then a bit longer than 2.59~\AA, and
HOWSs  show $d_{xz}$/$d_{yz}$ character accordingly.
When Au atom wire grows up to a certain length, 
the exchange energy gain thereby can split spin channels of $d_{xz}$/$d_{yz}$ orbitals,
and cause a crossover of stability of Au wire when $n=13$ as shown in Fig.\ref{fig02}(a).
As a result, the corresponding HOWSs change to $s$ character, in partial hybridization  with
$d_{z^2}$ orbitals. 

However, previous  reports indicate there is only one conduction channel 
around the Fermi level\cite{springborg07}, which suggests 
the emergence of 
 $d_{xz}/d_{yz}$  bands edges  around Fermi level for the
infinite wire are problematic and may account for the above crossover of
the cohesive energy.  
Moreover, strong localized $d_{x^2-y^2}$/$d_{xy}$ orbitals also 
mean that relevant self-interaction errors(SIE) cannot be ignored for 1-D wires\cite{wierzbowska05}.
We then employed GGA+U scheme to shift the \textit{d} bands and remove related SIE for verifications.
As shown in Fig.\ref{fig02}(d) and Fig.\ref{fig02}(e), even a small $U_{eff}$=1.0 eV can 
eliminate Stoner instability as no sharp DOS peaks emerge around the Fermi level, 
and relevant crossover at $n= 13$ atoms vanishes at the meantime. 
Correspondingly, the HOWSs of different wires bear  a similar characteristics with contribution of 
 hybridized states between $s$ and $d_{z^2}$ .
As displayed in Fig.\ref{fig02}(d),(h), HOWSs of 6-atom wire from GGA and GGA+U calculations are given, 
respectively. Obviously, GGA result shows a $d_{xz}$/$d_{yz}$ character, 
while GGA+U results of a $s+d_{z^2}$ hybridization orbital.
These results clearly indicate that improper descriptions on the $d$ bands 
may lead to the ferromagnetic behavior of the infinite wire and the crossover of the finite wires stability.
As revealed in Fig.\ref{fig02}(e), the cohesive energy  also follows the $a/x+b/x^2$ trend well.
Thus, the relativistic effects do not make much difference between stability trends of 
the Au and Ag wires but enhance binding interactions between Au atoms. 
This stronger binding property makes  quantum size effects of Au wires more easily 
perceptible in experiments, even under the influence of charge transfer or strain effect.
As reported in the experimental work by Crain \textit{et al.}, 
an even-odd oscillation exhibited in Au/Si(553) system at least up to 16 atoms\cite{crain06}.
Many relevant properties due to quantum size effects of Au atom wires 
are therefore both theoretically and experimentally predictible.

\subsection{ Ga, In and Pb Atom Wires}

Bulk Ga, In and Pb are typical \textit{p} electron metallic systems, 
and large efforts have been devoted to the formations of their corresponding 
atom wires in different vicinal surfaces to investigate  
the effect of \textit{p} electron
in 1-D systems\cite{snijders10,snijders05,gonzale04,albao05,javorsky09,kim07}. 
These atom wires showed some new behaviors in comparison with \textit{s}, or \textit{sd} systems.
Typically, the length distributions of Ga and In atom wires on Si(100) surface decrease 
monotonously for various coverages ,and show some  scale relations\cite{javorsky09,albao05}. 
Thus, it is of great interest to explore 
how the pure electronic contributions  affect  their length distributions.

\begin{figure}
	\includegraphics[width=0.72\textwidth]{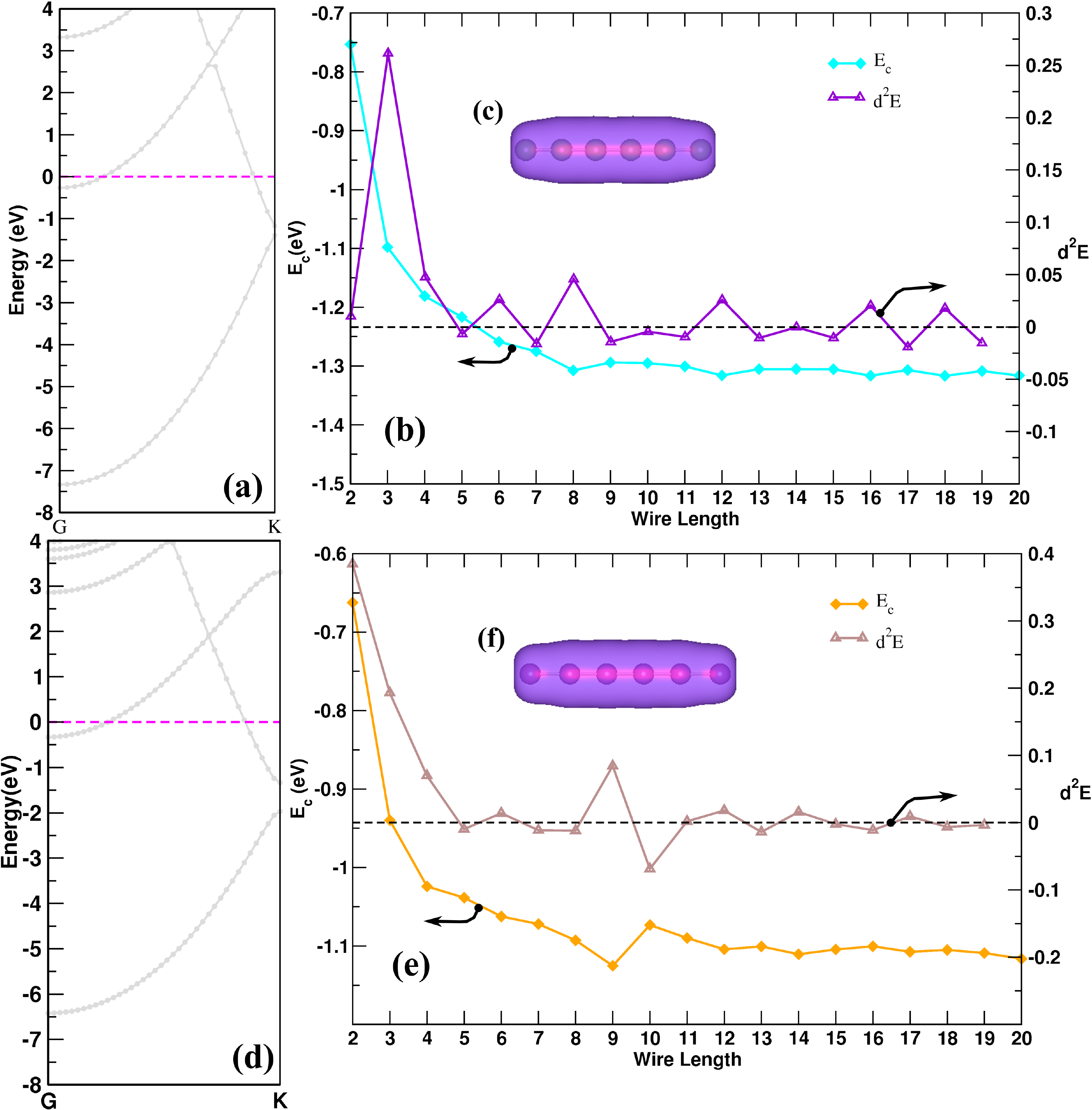}%
	\caption{(Color online)\label{fig03} (a) and (d), bandstructures of the infinite Ga atom wire and In atom wire,respectively, and the Fermi level is set to the zero energy. (b) is the cohesive energy $E_c$ of Ga atom wires versus length while (e) is that of In atom wires, and the second difference of $E_c$ is presented together. HOWSs for 6-atom Ga wire and In wire are given in (e) and (f),respectively. }
\end{figure}

We focus on Ga and In atom wires first. 
Both gallium and indium have 3 valences: 2 \textit{s} electrons and 1 \textit{p} electron 
in the outmost shell.  
Because of the large energy difference between the \textit{s} and \textit{p} electrons, 
little \textit{s-p} hybridization exhibits in these two elements 
as revealed by the bandstructures of the infinite wires  in Fig.\ref{fig03}(a) and (d),
and the stability of relevant atom wires is thus dominated by the \textit{p} electrons.
The bandstructures of Ga and In atom wires are quite similar, particularly for the valence bands, and both
have three band branches below the Fermi level.
Accordingly, the lowest branches are both of the \textit{s} character.
Then, the next one at $\Gamma$ point is predominated by $p_{x}$/$p_{y}$ electrons, 
and crosses over the Fermi level. The branch of $p_z$ character is also across the Fermi level
about $K$ point.
Hence, the combination of $p_z$ orbitals of atom wire forms $pp\sigma$ bonds, 
and $p_{x}$/$p_y$ orbitals contribute to $pp\pi$ states. 
This bonding property results in complicated behaviors of atom wires. 
As shown in Fig.\ref{fig03}(b) and (e), the cohesive energies of Ga and In atom wires 
both show oscillations beyond the even-odd pattern. 
Furthermore, the second difference of the cohesive energy $d^2E$
is also shown together, and exhibits a clear picture for the stability of atom wires.
It clearly shows that Ga atom wires of 2,3,4,6,8,12,16,18 atoms are stable segments,and In atom wires also
give a similar pattern and favor 2,3,4,6,9,12,14,17 atoms.
Since there is some uncertainty of 0.01 eV per atom to determine a local minimum for \textit{sp} systems,
some like 10-atom, 14-atom segments of Ga wires or 11-atom, 15-atom segments of In wires
probably lean to stable.
In other words, $d^2E$ of these wire segments is likely to be positive.
Therefore, $d^2E$ of Ga wires would display an even-odd oscillation around zero after 4-atom, 
which implies that 6-atom,8-atom,12-atom,16-atom, 18-atom, and probably 10-atom, 14-atom
are magically stable among others.
Meanwhile, $d^2E$ of In wires shows a bit compilcated oscillation around zero, 
and also exhibits some magical sizes such as 6-atom, 9-atom, 14-atom and 17-atom.
This fact would confound experimental observations out of the consideration of environmental factors like the charge transfer, which may lead to  different size distributions.
Although gallium and indium have the same valence configuration, a bit difference of  nuclei radius should 
result in this distinction in the stability. Actually, gallium and indium have different crystal structures 
due to the same fact.
6-atom HOWSs of Ga and In systems are then presented in Fig.\ref{fig03}(c) and (f),respectively, 
clearly showing a $pp\pi$ character. 

\begin{figure}
	\includegraphics[width=0.72\textwidth]{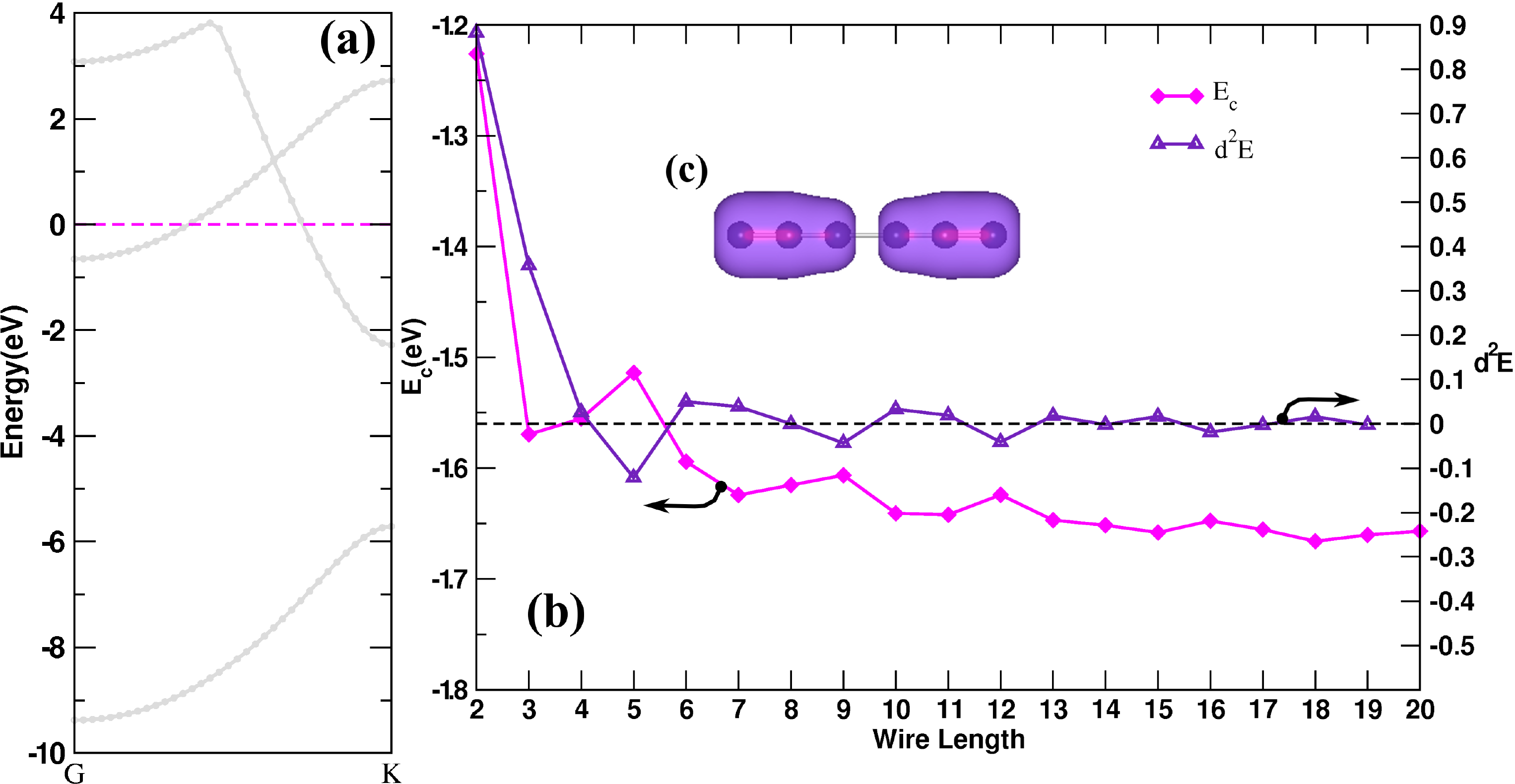}%
	\caption{(Color online)\label{fig04} (a), the bandstructure of the infinite Pb wire, and the Fermi level is set to the zero energy; (b) The cohesive energy $E_c$ of Pb atom wires versus lengths, together with the second difference of $E_c$, (c) is HOWS of 6-atom Pb wire.   }
\end{figure}

Like gallium and indium, lead also has little \textit{sp} hybridization 
when forming the atom wires, 
which leads to only two \textit{p} electrons contributing to 
binding interactions as revealed by 
the bandstructure of infinite wire shown in Fig.\ref{fig04}(a). 
The valence bands are also similar to that of Ga wire or In wire, with three band branches below
the Fermi level. 
The \textit{s} band, the lowest branch, is shown about 3.5 eV from $p_z$ branch at $K$ point, 
while both $p_x$/$p_y$ and $p_z$ branches cross over the Fermi level in the middle 
of 1-D Brillouin zone.
Due to axial symmetry, $p_{x/y}$ orbitals 
are two-fold degenerate and contribute to $pp\pi$ states of atom wire
while the combination of $p_z$ orbitals forms $pp\sigma$ bonds. 
Such binding behaviors are quite similar to those of Ga and In atom wires,
but involve  2 \textit{p} electrons per atom to be filled up. 
Therefore, the stability of Pb atom wires should present a certain 
oscillation beyond the even-odd feature as displayed in Fig.\ref{fig04}(b). 
And based on the second difference $d^2E$ in  Fig.\ref{fig04}(b),
we find the stable systems are of 2,3,4,6,7,10,11,13,15,18 atoms, respectively.
Meanwhile, relative large amplitude of oscillation also suggests 
this stability trend survives even under environmental influences, 
within some preferred lengths in fabrications.
As for a Pb dimer, two \textit{p} electrons occupy $pp\sigma$ bond 
and other two then go into $pp\pi$ bond. Interestingly, 
HOWS of Pb dimer is $pp\sigma$ bonding state while that of other lengths 
is relevant $pp\pi$ state. 
Concerned with 3-atom wire, two electrons go into $pp\sigma$ bond, 
and the rest 4 \textit{p} electrons then fully fill up the two-fold degenerate $pp\pi$ bonding state. 
The full occupation of bonding states makes 3-atom wire inert, and
explains why 3-atom wire is stable among others.
Thus, when the atom wire becomes longer, more complicated electronic states are required to fill up, contributing to this specific behavior. 
At the end, HOWS of 6-atom Pb wire is then given in Fig.\ref{fig04}(c), showing a typical $pp\pi$ character
with a node in the middle of the atom wire.

Results on free-standing atom wires definitely oversimplify the stability trends of 1-D atom wires.
The adsorption of atoms on metallic or semiconducting substrates should lead to certain 
hybridization of atom orbitals with  electronic states of substrates.  
In fact, the interactions between single atoms in 1-D wires  
 result from a direct overlap between atoms' wavefunctions and substrate-mediated mechanisms, 
such as the interface Friedel oscillations\cite{zz98}.
Thus, the tradeoff between these two interactions would result in different length distributions in 
experimental observations. As for the  Au atom wires, 
the strong \textit{sd} hybridization enhances binding interaction between Au atoms,
leading to a pronounced even-odd oscillation of stability even for long wires. 
Consequently, relevant quantum size effect was observed in Au/Si(553) system 
while few work touches this issue for Na or Ag wires\cite{crain06}.  
This behavior implies  atom wires with relative large amplitude of oscillations 
like Ga and Pb  are likely to have quantum size phenomenon even under the influence of substrate or strain effect. 
Therefore, more works are expected to explore relevant quantum size effects 
in atom wires like Ga and Pb wires.



\section{CONCLUSION}

In conclusion, we have given a systematic study on electronic stability 
and quantum size effect of 1-D metal atom wires by extensive \textit{ab initio} calculations. 
The results show that the cohesive energy of Na atom wires presents a typical even-odd oscillation 
with a $a/x+b/x^2$ trend, which can be ascribed to  a typical 1-D quantum confinement 
as well as the pair-up of electrons.
Meanwhile, Ag atom wires show a similar behavior but with a stronger binding interaction 
due to certain \textit{sd} hybridization.  
A good agreement with the $a/x+b/x^2$  trend actually suggests the hybridized states of Ag are still 
quite delocalized and undirectional.
As for Au atom wires, short ones that are less than 13 atoms show a similar even-odd oscillation. 
Once the length increases to 13 atoms, a crossover occurs. 
Such behavior is due to the emergence of $d_{xz}/d_{yz}$ states into bonding  interactions, 
which also results in  ferromagnetic property of the infinite wire 
caused by 1-D von Hove singularities around Fermi level. Further GGA+U calculations show that even a small $U_{eff}$ =1.0 eV can eliminate the relevant crossover 
when $n=13$ and the  ferromagnetic behavior of the infinite Au wire, 
which implies a proper description of the \textit{d} states can affect significantly the behaviors of 
the Au wires.
The large \textit{sd} hybridization due to the relativistic effect enhances bindings between 
the Au atoms, relevant even-odd oscillation thereby can be observed experimentally.
With respect to \textit{sp} systems examined here, 
we find the stabilities of Ga, In and Pb atom wires are all dominated
by binding interactions between the \textit{p} electrons because of little hybridization 
between the \textit{s} and \textit{p} electrons.
Thus, the directional binding behaviors of the \textit{p} electrons, e.g.,
$pp\sigma$ due to the combination of $p_z$ orbitals and 
$pp\pi$ from $p_x/p_y$ orbitals  result in intricate oscillations in the stability trends, 
and suggest relevant status of electrons filling-up determines the dominant behavior.  

\begin{acknowledgments}
 This work was supported in part by NSF(Grant No. DMR-0906025). 
 The calculations were performed at NERSC of DOE.
\end{acknowledgments}





\clearpage

\end{document}